\documentstyle[12pt]{article}

\def\a{\alpha}
\def\b{\beta}

\def\d{\delta}
\def\e{\epsilon}                

\def\h{\eta}

\def\l{\lambda}
\def\m{\mu}
\def\n{\nu}

\def\p{\pi}                     
\def\r{\rho}                    
\def\s{\sigma}                  

\def\eq{\begin{equation}}
\def\eqe{\end{equation}}
\def\eqa{\begin{eqnarray}}
\def\eqae{\end{eqnarray}}

\newcommand{\half}{\frac{1}{2}}

\newcommand{\del}{\partial}

\catcode`\@=11
\def\@afterindentfalse{\let\if@afterindent=\iftrue}
\catcode`\@=12

\def\del{\partial}
\def\half{\frac{1}{2}}

\def\pl{Phys. Lett.\ }

\def\np{Nucl. Phys.\ }
\def\cin{c\rightarrow \infty}

\def\dihma{\d^\infty h_\m{}^A}
\font\biggbold=cmbx10 scaled\magstep2

\font\bigreg=cmr10 at 12pt
\bigreg
\begin{document}
\baselineskip =15pt

\vskip 2cm

\hspace{4.0in} ITP-SB-92-67

\vskip .1cm

\hspace{4.0in} LBL-33339

\vskip .1cm

\hspace{4.0in} UCB-PTH-92/43

\vskip 2cm

\begin{center}
{\biggbold{\Large {\bf Quantum gauging from classical gauging}}}\\
{\biggbold{\Large {\bf of nonlinear algebras}}}
\end{center}

\vskip 2cm

\centerline{\bf A. Sevrin%
  \footnote{Theoretical Physics Group, Lawrence Berkeley Laboratory, 1
Cyclotron Rd, and Department of Physics, University of California at
Berkeley, Berkeley, CA 94720, USA} and P. van Nieuwenhuizen%
  \footnote{Institute for Theoretical Physics, State University of New York
at Stony Brook, Stony Brook, NY 11794-3840, USA}}

\vskip 2cm

\noindent {\bf Abstract:} We extend our previous theory of the gauging of
classical quadratically nonlinear algebras without a central charge but with
a coset structure, to the quantum level. Inserting the minimal anomalies into
the classical transformation rules of the currents introduces further quantum
corrections to the classical transformation rules of the gauge fields and
currents which additively renormalize the structure constants.  The
corresponding Ward identities are the $c \rightarrow \infty$ limit of the full
quantum Ward identities, and reveal that the $c \rightarrow
\infty$ limit of the quantum gauge algebra closes on fields and currents.
Two examples are given.

\newpage
\baselineskip=24pt

\section{Introduction and summary.}
The past two years several new gauge theories have been constructed
\cite{hull1,us1,us3,delius}, which are extensions of two-dimensional gravity
\cite{pol1} and which
are based on so-called nonlinear Lie algebras \cite{zamo,uscmp1}.  The latter
are algebras
of the kind $[T_A, T_B]=T_C f^C{}_{AB} + T_C T_D V^{DC}{}_{AB} + \ldots$.
We shall
restrict our attention to quadratically nonlinear algebras.  To every
generator $T_A$
one associates a gauge field $h_\m{}^A$ and a classical current $u_A{}^\m$.
   To
construct corresponding quantum currents, one proceeds as follows.  One
begins by
coupling matter to the external gauge fields $h_\m{}^A$ and fixing  all local
symmetries without anomalies. By integrating out the matter, one finds the
induced
action $S^{ind}$ which depends on the gauge fields and on one (or more) central
charge $c$. This central charge characterizes the matter system and is
for example the number of scalar fields coupling to the gauge fields. Defining
the quantum currents as $u_A{}^\m \sim \del S^{ind} /\del h_\m{}^A$, one
can then derive Ward identities for the remaining anomalous symmetries. These
Ward identities provide functional differential equations for the induced
action.
They are of the generic form $D_\m u_A{}^\m = An_A$, where $D_\m$ is a
covariant derivative and $An_A$ are the anomalies.  The anomalies $An_A$
are as always local expressions depending on $h_\m{}^A$, but the expression
$D_\m u_A{}^\m$ is in general nonlocal and very complicated.  However, for
$c \rightarrow \infty$ it reduces to a local functional of $h_\m{}^A$ and
$u_A{}^\m$.  Namely, in addition to terms linear in $h_\m{}^A$, the covariant
derivatives $D_\m$ also contain terms bilinear in $h_\m{}^A$ and
$u_A{}^\m$, which are due to the nonlinear terms in the algebra.

{}From this $c \rightarrow \infty$ Ward identity one can extract the
$c\rightarrow \infty$ transformation laws of the quantum currents,
denoted by $\d^\infty u_A{}^\m$, and by requiring invariance of the $c
\rightarrow \infty$ Ward identity, one can deduce the corresponding
$c\rightarrow \infty$ transformation laws of the gauge fields, denoted by
$\d^\infty h_\m{}^A$.  In examples \cite{us3,delius} it has been found that
these
$\d^\infty h_\m{}^A$ and $\d^\infty u_A{}^\m$ form a closed gauge algebra.
\footnote{Since all symmetries without anomalies have been fixed, one does
not have gauge fixing terms and corresponding ghosts; consequently, one
deals with gauge transformations and not BRST symmetries.}

In this article we shall consider nonlinear Lie algebras defined by an
operator product expansion (OPE).  These algebras are quantum algebras
with central charges.  To obtain a
corresponding classical nonlinear Lie algebra without central charges we
first take the $c \rightarrow \infty$ limit of the full
nonlinear quantum algebra (which converts the quantum algebra to a
Poisson bracket algebra but with central charge) and then take in the
result the $c \rightarrow 0$ limit (which removes the central charge).
The theory of the gauging of such classical nonlinear Lie algebras was
established by us in collaboration with K. Schoutens \cite{us5,usym}, and it
leads to classical
$\d_{cl} h_\m^A$ and $\d_{cl}u_A{}^\m$ with an {\bf open} classical
nonlinear gauge algebra. Using this formalism as a starting point, we shall
construct the $\d^\infty u_A{}^\m$ by adding the minimal anomalies to
$\d_{cl} u_A{}^\m$ and requiring closure of the algebra of $\d^\infty
u_A{}^\m$.  In this way we find that the structure constructs $f^A{}_{BC}$
become replaced by $f^A{}_{BC} + g^A{}_{BC}$ where $g^A{}_{BC}$ can be
viewed as additive renormalizations.  From these $\d^\infty u_A{}^\m$ one
can also (re)construct the $c\rightarrow \infty$ Ward identities.  In the
original classical gauge algebra without central charges, closure of the gauge
algebra on $h_\m{}^A$ is violated by terms proportional to $D_\m{}^{cl}
u_A{}^\m$.  In the new $c\rightarrow \infty$ gauge algebra with central
charges, the extra terms in $\d^\infty u_A{}^\m$ due to the anomalies
complete $D_\m{}^{cl} u_A{}^\m$ to the full $\cin$ Ward identity and as a
result, the algebra of $\d^\infty h_\m{}^A$ now also closes. Hence, the gauge
algebra becomes uniformly closed due to quantum effects.  \footnote{For
those induced gauge theories which can be obtained from WZWN models
(with a suitable gauge group) by imposing constraints on the WZWN currents
\cite{us3}, closure of the $\d^\infty$ algebra is guaranteed, since the local
gauge parameters in the induced gauge theories are obtained from those of
the WZWN models by requiring the constraints to be preserved by the local
symmetries of the induced gauge theories.}

The idea that the anomalies
introduce extra terms in $\d_{cl} u_A{}^\m$ which close the gauge algebra of
$\dihma$, was first proposed in \cite{delius}.  However, as we have found, the
extra terms are not only the anomaly itself, but also terms which are of the
same form as the term $u_C{}^\m f^C{}_{AB}\e^B$.  These extra terms can be
written as $u_C{}^\m g^C{}_{AB} \e^B$ and renormalize the structure
constants as mentioned above.

Our results give a very simple algorithm to construct the $c \rightarrow
\infty$ quantum algebras from the corresponding (usually much simpler)
classical algebras.  Input is the existence of the $c \rightarrow \infty$
algebra, output all (often complicated) quantum corrections in explicit form.

In section 2 we review the gauging of classical nonlinear algebras, and give
the example of chiral $W_3$ gravity.  In section 3 we extend these results
to the quantum level, and work through the  cases of $W_3$ gravity and
chiral supergravity.  In section 4 we discuss to which OPE our new
formalism can be applied, and we comment on the relations to our earlier
work.

We should perhaps stress that we have no corresponding results for finite
$c$. However, when passing from the induced to the effective theory ({\it
i.e.}, integrating over the gauge fields), one finds that the Ward identities
for the effective theory at arbitrary values of $c$ are the same, up to
multiplicative renormalizations, as those for the induced theory for
$c\rightarrow\infty$\cite{us6}. For a discussion of the quantum theory for
general $c$
and its
renormalizability, see \cite{us6,deb1}.

\section{Gauging of classical nonlinear algebras without central charge.}
In \cite{us5} we developed a theory of classical gauging of nonlinear
algebras.
The simplest case are the quadratically nonlinear algebras, defined by
\eq
[T_A, T_B ] = m_{AB} + T_C f^C{}_{AB} + T_D T_C V^{CD}{}_{AB}
\label{one}
\eqe
where $m_{AB}$ are the central charges.  When dealing with classical
algebras, the brackets may be realized by Poisson brackets, and no ordering
ambiguities in the last term arise. However, if the $T_A$ are quantum
operators, one must normal-order the last term on the right-hand side.  The
Jacobi identities restrict the constants $m_{AB}, f^A{}_{BC}$ and
$V^{CD}{}_{AB}$, and those for the quantum algebras contain extra terms due
to the nonassociativity of the normal-ordered product \cite{uscmp1}.

An important subclass are the nonlinear algebras with a coset structure. The
generators $T_A$ can then be divided into coset generators $K_\a$ and
generators $H_i$ which form a subalgebra, such that only $V^{ij}{}_{\a\b}$ is
nonvanishing. In fact, all known quadratically nonlinear Lie algebras are of
this kind.  For these algebras a nilpotent BRST charge $Q$ has been written
down, both at the classical level and at the quantum level \cite{uscmp1}
\footnote{The quantum $Q$ is obtained from the classical $Q$ by a
renormalization of the $f^A{}_{BC}$ proportional to $V^{AB}{}_{CD}
f^D{}_{BE}$.}. For the classical quadratically nonlinear Lie algebras with
coset
structure, the Jacobi identities read
\eqa
m_{AD} f^D{}_{BC} &+& ({\rm cyclic \; in} \; ABC) = 0 \nonumber\\
f^E{}_{AD} f^D{}_{BC} &+& 2 m_{AD} V^{DE}{}_{BC} + ({\rm cyclic \; in} \; ABC)
=
0
\nonumber\\
V^{EF}{}_{AD}  f^D{}_{BC} &+& f^E{}_{AD} V^{DF}{}_{BC} + f^F{}_{AD}
V^{DE}{}_{BC} + ({\rm cyclic \; in} \; ABC) = 0
\label{oneA}
\eqae

We consider now classical quadratically nonlinear algebras with a coset
structure, but without central charges.  One associates with every $T_A$ a
gauge field $h_\m{}^A$ and a current $u_A$.  Since in all cases considered
the index $\m$ of the gauge fields and currents takes only one value (for
example, $h_{++}$ or $u_{--}$) we shall suppress the index $\m$ of
$u_A{}^\m$, but keep it on $h_\m{}^A$ for historical reasons.  These gauge
fields and currents transform classically as follows
\eqa \d_{cl} h_\m{}^A
&=& \del_\m \e^A + (f^A{}_{BC}  + 2 u_D V^{DA}{}_{BC}) h_\m{}^C
\e^B \nonumber\\
\d_{cl} u_A &=& u_C (f^C{}_{AB} + u_D V^{DC}{}_{AB}) \e^B \label{two}
\eqae
We shall presently comment on the factor 2. The gauge commutator closes on
$u_A$  \eq
[\d_{cl} (\e_1), \d_{cl} (\e_2) ] u_A = \d_{cl} (\e_3) u_A
\label{three}
\eqe
where the structure constants are source-dependent
\eq
\e_3{}^A = \tilde{f}^A{}_{BC} \e_1{}^C \e_2{}^B, \tilde{f}^A{}_{BC} =
f^A{}_{BC}
+ 2 u_D V^{DA}{}_{BC}
\label{four}
\eqe
The covariant derivatives of the sources follow from (3)
\eq
D_\m{}^{cl} u_A = \del_\m u_A - u_C f^C{}_{AB} h_\m{}^B -  u_D u_C
V^{CD}{}_{AB} h_\m{}^B
\label{five}
\eqe
and transform in the coadjoint representation
\eq
\d (D_\m{}^{cl}  u_A) = (D_\m{}^{cl} u_C) \tilde{f}^C{}_{AB} \e^B
\label{six}
\eqe
The gauge commutator on $h_\m{}^A$ does not close; it is off by a term
proportional to $D_\m{}^{cl} u_D$
\eq
[\d (\e_1), \d (\e_2) ] h_\m{}^A = \d (\e_3) h_\m{}^A - 2 D_\m{}^{cl} u_D
V^{DA}{}_{BC} \e_1{}^C \e_2{}^B
\label{seven}
\eqe
Curvatures are defined by
\eq
[D_\m , D_\n ] u_A = - u_C (f^C{}_{AB} +  u_D V^{DC}{}_{AB}) R_{\m\n}{}^B
\label{eight}
\eqe
and $R_{\m\n}{}^A$ is given by the usual expression, but with
$\tilde{f}^A{}_{BC}$ instead of $f^A{}_{BC}$.  The curvatures transform in the
adjoint representation plus a term proportional to the covariant derivative
of the sources
\eq
\d R_{\m\n}{}^A = \tilde{f}^A{}_{BC} R_{\m\n}{}^C \e^B + \left\{ 2 D_\m{}^{cl}
u_D  V^{DA}{}_{BC} h_\n{}^C \e^B - \m \leftrightarrow \n \right\}
\label{nine}
\eqe
The covariant derivatives of the curvature tensor are given by
\eq
D_\r{}^{cl}  R_{\m\n}{}^A = \del_\r R_{\m\n}{}^A - \tilde{f}^A{}_{BC}
R_{\m\n}{}^{C} h_\r{}^B -  (D_\m{}^{cl}  u_D V^{DA}{}_{BC} h_\n{}^C h_\r{}^B -
\m \leftrightarrow \n )  \label{ten}
\eqe
Note that in (11) the factor 2 which appeared in (10), has been replaced by
unity.  They satisfy the Bianchi identities
\eq
D_{[\m}{}^{cl}  R_{\n\r]}{}^A = 0.
\label{eleven}
\eqe

The previous results require only the Jacobi identities in (2) with
$m_{AB}=0$. Since the latter remain valid upon scaling $V^{AB}{}_{CD}$ by
a factor, the results in (3) contain  a free parameter.  As we will see in the
next section,  inclusion of central charges fixes the scale of
$V^{AB}{}_{CD}$,
and the correct results are those given in (3).  In our previous work on
classical $W$ gravity we had chosen the normalization
$\d_{cl} h_\m{}^A  =  \ldots + u_D V^{DC}{}_{AB} \ldots$ and
$\d_{cl} u_A  =  \ldots + \half u_D V^{DC}{}_{AB} \ldots$  which
differs from the present rules by a factor 2.

The classical nonlinear gauge algebras with $c=0$  discussed in this section
arise as symmetries of the classical action for the coupling of matter to
external gauge fields.  A typical example is chiral $W_3$ gravity.  The
classical matter-coupled action reads [1]
\eqa
S_{cl} &=& \frac{1}{\p} \int \big[ - \half \del_+ \varphi^i \del_- \varphi^i -
h_{++} T_{--} - b_{+++} W_{---} \big] d^2 x \nonumber\\
T_{--} &=&- \half \del_- \varphi^i \del_- \varphi^i \; , \; W_{---} = -
\frac{1}{3} d_{ijk} \del_- \varphi^i \del_- \varphi^j \del_- \varphi^k
\eqae
with $d_{ijk}$ a totally symmetric constant symbol satisfying $d_{i(jk}
d_{\ell)mi} = \d_{i(jk} \d_{\ell)mi}$.  The symmetries are given by
\eqa
\d \varphi^i &=& \e \del_- \varphi^i + \l_{++} \del_- \varphi^j \del_-
\varphi^k d_{ijk} \nonumber\\
\d h &=& \del_+ \e - h \del_- \e + \e \del_- h + (\l \del_- b - b \del_- \l)(-
2 T_{--} ) \nonumber\\
\d b &=& \e \del_- b - 2 b \del_- \e + \del_+ \l - h \del_- \l + 2 \l \del_- h
\eqae
The local gauge algebra reads [13]
\eqa
\left[ \d (\e_1) , \d (\e_2) \right] &=& \d \left( \e_3 = - \e_1 \del_- \e_2 +
\e_2 \del_- \e_1 \right) \nonumber\\
\left[ \d (\e) , \d (\l) \right] &=& \d \left( \hat{\l} = - \e \del_- \l + 2
\l
\del_- \e \right) \nonumber\\
\left[ \d (\l_1), \d (\l_2) \right] \left(\begin{array}{c} \varphi^i \\ h \\
b \end{array} \right) &=& \d (\hat{\e}) \left( \begin{array}{c} \varphi^i \\ h
\\ b \end{array} \right) + \left( \begin{array}{l} 2 \p k \del_- \varphi^i \d
S/\d h \\ - 2 \p k \del_- \varphi^i \d S/\d \varphi^i  \\ \qquad 0 \end{array}
\right) \nonumber\\
\hat{\e} &=& - 2 k T_{--} \; , \; k = - \l_1 \del_- \l_2 + \l_2 \del_- \l_1
\eqae
Hence on $\varphi^i$ it is closed up to the $h$ field equation, and on $h$ up
to the $\varphi$ field equation.  The terms with field equations form
separately a trivial symmetry, which explains the relative minus sign and
the common factor $(2\p k \del_- \varphi^i)$.

Defining $u = -\frac{1}{2} (\del \varphi)^2$ and $v = -\frac{1}{3} d_{ijk}
\del
\varphi^i \del \varphi^j \del \varphi^k$  with $\del = \del_-$ one finds
\eqa
\d_{cl} u &=& 2 \del \e u + \e \del u + 2 \l \del v + 3 \del \l v \nonumber\\
\d_{cl} v &=& 3 \del \e v + \e \del v -4 (\del \l u u + \l u \del u )
\eqae
In order that this corresponds to (3), the antisymmetry of $f^A{}_{\a i}$ in
$\a, i$ requires that $\int [\l \d(\e)v+\e \d (\l) u ] d^2 x$ vanishes.  This
fixes the scale (and the form) of the $\d (\l) u$ terms with respect to the
$\d
(\e) v$ terms (which accounts for the absence of a factor $\frac{1}{15}$
which is present in our earlier work [3]).   One easily verifies that these
currents
satisfy the same classical closed gauge algebra as the gauge field $b$,
except that $\hat{\e}$ is now twice as large
\eq
\left[ \d_{cl} (\l_1 ), \d_{cl} (\l_2) \right]  \left( \begin{array}{c}u\\v
\end{array}
\right) = \d_{cl} (\hat{\e} = -4 (\l_2 \del \l_1 - \l_1 \del \l_2 ) u )
\left( \begin{array}{c}u\\v \end{array} \right)
\eqe
In section 4 we shall explain this difference by a factor 2, but now we
continue with (17).

In our earlier work we stated that the classical gauge
algebra closes on the gauge fields up to field equations.  Although this is
correct, from our present perspective we prefer to interpret the term
$-2 \p k \del_- \varphi^i  \del S/\del \varphi^i$ in (15) as a covariant
derivative
of the currents, namely as $-D_\m{}^{cl}  u_D V^{DA}{}_{BC} \l_1{}^C \l_2{}^B$
with $u_D = u, V^{DA}{}_{BC} \l_1{}^C \l_2{}^B = - 2(\l_2 \l_1^\prime - \l_1
\l_2^\prime )$ and $D_\m{}^{cl} u_D$ given by
\eq
D_+{}^{cl} u = (\del_+ - 2 \del_- h  - h \del_- ) u - (3\del_-  b + 2
b \del_- ) v
\eqe
The transformation rules $\d_{cl} h_\m{}^A$ in (3),which correspond to the
gauging of
the algebra, are obtained from the transformation rules in (14), which leave
the
classical action invariant,by rescaling the $T$ term in (14) by a factor 2,
for reasons
to be explained in section 4.  The gauge algebra for $h, b, u$ and $v$
(with $\d_{cl} (\l)
h = (\l \del_- b_- b \del_- \l) (-4 u))$  is then an example of a classical
nonlinear
gauge algebra with $c=0$.

\section{Gauging at the quantum level.}
To extend the gauging of the classical theory to the quantum level, we must
integrate out the matter. This will introduce an anomaly which shows up as a
violation of current conservation, described by a Ward identity of the form
$\del_+ u_A = (An)_A +$ more.  The minimal anomaly is due to the variation of
the two-point function with two external gauge fields, under the leading
variation $\d h_\m{}^A = \del_\m \e^A$.  By varying the Ward identity, we
find the minimal anomaly in the transformation rules of the currents
\eq
\d_{min} u_A = m_{AB} \e^B
\eqe
For example, in ordinary gravity, the anomaly appears as
$\del_+ u = \del_-{}^3 h +  \ldots$, and in $W_3$ gravity this result is
extended to the spin 3 current $v$ as $\del_+ v = \del_-{}^5 b + \ldots$.
Then (19) corresponds to $\d u = \del_-{}^3 \e + \ldots$ and $\d v =
\del_-{}^5 \l + \ldots$.

The classical transformation rules of the currents plus the terms due to the
minimal anomaly are only part of the $c \rightarrow \infty$
transformation rules
\eq
\d^\infty u_A = u_C (f^C{}_{AB} + u_D V^{DC}{}_{AB} ) \e^B + m_{AB} \e^B +
\ldots
\eqe
We shall now make the following assumptions (to be discussed later)
concerning the terms denoted by dots in (20)
\begin{itemize}
\item [(i)] the $u_i$ transform under $\e^A$ as in the classical theory
(except for minimal anomalies)
$$ \d^\infty (\e^A) u_i = m_{iA} \e^A+ \d_{cl} (\e^A) u_i $$
\item [(ii)] the $u_\a$ transform under $\e^i$ as in the classical theory.
This follows
actually from the antisymmetry of the structure constants and (i), if the
general
result in (23) is to hold.
\item [(iii)] the minimal anomalies respect the coset structure
$$ m_{\a i} = 0 $$
In fact, in all applications $m_{AB} \e^B$ will always by proportional to
$\del^m \e^A$ with $m$ a positive integer.
\item [(iv)] the $\d^\infty$ gauge algebra closes.
\end{itemize}
In section 4 we shall discuss which quantum algebras fulfill these
assumptions.
 We can then compute the composite parameters $\e_3{}^A$ of
the $\d^\infty$ gauge algebra by evaluating the gauge commutator on
$u_i$, keeping only the field independent terms (which are due to the
minimal anomaly) and rewriting them as the minimal anomaly in terms of
$\e_3{}^A$. In formula
\eqa
\left[ \d (\e_1{}^A) , \d (\e_2{}^A) \right] u_i &=& m_{CD} \e_1{}^D
f^C{}_{iB}
\e_2{}^B -  1 \leftrightarrow 2 \nonumber\\
&=& m_{iB} \e_3{}^B
\eqae
The composite parameters $\e_3$ is parametrized as follows:
\eqa
\e_3{}^A &=& f^A{}_{BC} \e_1{}^C \e_2{}^B + g^A{}_{BC} \e_1{}^C \e_2{}^B
\nonumber\\
&+& 2 u_D V^{DA}{}_{BC} \e_1{}^C \e_2{}^B
\eqae
The $f^A{}_{BC}$ are the structure constants of the classical theory but the
$g^A{}_{BC}$ are determined by this expression and are the corrections
induced by the minimal anomaly.  Given the minimal anomaly $m_{AB} \e^B$
with a given overall scale, the Jacobi identity in (2) fixes the scale of
$V^{AB}{}_{CD}$. This results in a factor unity in front of the $V$ term in
$\d u_A$
\eq
\d^\infty u_A = m_{AB} \e^B + u_C (f^C{}_{AB} + g^C{}_{AB} + u_D
V^{DC}{}_{AB} ) \e^B
\eqe
The claim is that with these transformation rules, the gauge algebra closes
on all currents $u_A$.  To prove this, one needs all three Jacobi
identities in (2).

It may be helpful to give an example at this point.  The $c \rightarrow
\infty$ limit of the full quantum transformation rules of the currents of
$W_3$ gravity is given by \cite{us3}
\eqa
\d^\infty u  &=& -4 \del^3 \e + \e \del u + 2 u \del \e + 2 \l \del v + 3
\del \l
v \nonumber\\
\d^\infty v &=& -4 \del^5 \l+ \e \del v + 3 \del \e v - 4 (\del \l uu + \l u
\del u
) \nonumber\\
&+& [2 \l \del_-{}^3 + 9 \l^\prime \del_-{}^2 + 15 \l^{\prime\prime} \del_-
+10 \l^{\prime \prime \prime} ] u
\eqae
We shall reobtain them from the corresponding classical transformation
laws by our method.  These classical transformation laws are obtained by
dropping the minimal anomalies ($\del^3\e$ and $\del^5 \l)$ as well as the
terms in square brackets.  The latter correspond to the terms due to
$g^A{}_{BC}$, but we pretend at this point that we do not know them.  (In
section 4 we shall show that in general the $g^A{}_{BC}$ terms in $\d u_\a$
correspond to all terms linear in currents and proportional to $\e^\b$).

We begin our program by adding the minimal anomalies, which we parametrize as
$\a \del^3 \e$ and $\b \del^5 \l$ with $\a$ and $\b$ constants to be
determined.  Next
we consider second variations of $\d^\infty u$.  Under two variations with
$\e^i
\equiv \e$ we find, collecting all field-independent terms,
\eqa
\left[ \d^\infty (\e_1), \d^\infty (\e_2) \right] u &=& \a \e_2 \del (\del^3
\e_1) + 2 (\del^3 \e_1) \del \e_2 - 1 \leftrightarrow 2 + \ldots \nonumber\\
&=& \a \del^3 (- \e_1 \del \e_2 + \e_2 \del \e_1) + \ldots
\eqae
This is the same result as for $\d_{cl} (\e)$ in (15), and shows that
$g^A{}_{ij} =
0$.
Next we evaluate the commutator of $\d (\e_i)$ and $\d (\e_\a)$, keeping
only the terms linear in currents
\eqa
&& \left[ \d^\infty (\e), \d^\infty (\l) \right] u  = (2 \l \del + 3
\l^\prime) ( \e
v^\prime + 3 \e^\prime v ) \nonumber\\
&& - (\e  \del + 2 \e^\prime )(2\l v^\prime + 3 \l^\prime v) = (2 \tilde{\l}
\del + 3 \tilde{\l}^\prime ) v = \d^\infty (\tilde{\l})v
\eqae
with
\eq
\tilde{\l} = - \e \del \l + 2 \l \del \e
\eqe
Again this is the same result as for $\d_{cl}$ in (15), and shows that also
$g^A{}_{\a
j} = 0$.  However, for the commutator of two coset transformations $\d
(\e_{1 \a})$ and $\d (\e_{2\a})$ on $u_i$ we find a nontrivial result.
Retaining again only the field-independent terms, we find
\eqa
&& \left[ \d (\l_1), \d (\l_2) \right] u = \b  (2 \l_2 \del + 3 \l_2^\prime
)(\del^5
\l_1) - 1 \leftrightarrow 2 + \ldots \nonumber\\
&& = \b \del^3 (-2 \l_1  \l_2^{\prime\prime\prime} + 3 \l_1^\prime
\l_2^{\prime\prime} - 3 \l_1^{\prime\prime} \l_2^\prime + 2
\l_1^{\prime\prime\prime} \l_2 ) + \ldots \nonumber\\
&& = \b \del^3 ( \tilde{\e} )+ \ldots
\eqae
Since classically there is no field-independent contribution to this
commutator, $f^i{}_{\a\b} = 0$ and hence the present $\tilde{\e}$ is
completely due to a new structure constant $g^i{}_{\a\b}$. This
$g^i{}_{\a\b}$ contributes a term $\d v_\a = u_i g^i{}_{\a\b} \e^\b$, given by
\eq
\d v = (10 \l^{\prime\prime\prime} + 15 \l^{\prime\prime} \del_- + 9
\l^\prime \del_-{}^2 + 2 \l \del_-{}^3 ) u
\eqe
(It follows most easily by extracting $\l_2$ from $\int u g^i{}_{\a\b}
\l_1{}^\b \l_2{}^\a d^2 x)$.  This is precisely the term in square brackets in
$\d^\infty v$ in (24).  To complete the transformation rules of the $W_3$
currents,
we evaluate $[\d (\e), \d (\l)] v$ which yields $\a = \b = - 4$.  These
results agree
with (24).

As a second example, consider induced (3,0) supergravity \cite{delius}.  The
OPE has one central charge, $\s$.  The transformation rules for the currents
in the limit
$\s \rightarrow \infty$ are given by
\eqa
\d^\infty u &=& \e u^\prime + 2 \e^\prime u + \l_a^\prime v^a + \frac{3}{2}
\h_i^\prime q^i + \frac{1}{2} \h^i q_i^\prime \; [+\del_-{}^3 \e ] \nonumber\\
\d^\infty v_a &=& \e v_a^\prime + \e^\prime v^a - \frac{i}{2} \e^{abc} \l_b
v_c - \frac{i}{2}
\e^{aij} \h_i q_j \;  [+ \del_- \l^a ] \nonumber\\
\d^\infty q_i &=& \e q_i^\prime + \frac{3}{2} \e^\prime q_i - \frac{i}{2}
\e_{iaj} \l^a
q^j -\frac{1}{4}  \h^j v_j v_i  \nonumber\\
&& [ + \half \h_i{} u - \frac{i}{2} \e_{ija} (2\h_j^\prime + \h_j \del) v^a +
\del_-{}^2 \h_i ]
\eqae
The parameters $\e$ and $\l^a (a=1,3)$ correspond to the stress tensor $u$
and the $SO(3)$ current $v_a$, while $\h^i (i=1,3)$ correspond to the
supersymmetry
currents $q_i$.  The currents $u$ and $v_a$ constitute the subalgebra
currents $u_i$,
and $q_i$ corresponds to the coset currents $u_\a$, but in $\d q_i$ there
is only a
nonlinear term with $vv$ but no terms with $vu$ or $uu$.  (Actually, we can
also
consider $u$ as a coset current.  The coset currents $u_\a$ are then $q_i$
and $u$,
while $v_a$ are then the only subalgebra currents $u_i$.  It will be shown
in section 4
that this division does not lead to a corresponding classical algebra.  In
general, {\bf
only} the $u_A$ for which $\d u_A$ contains nonlinear terms are to be
considered as
coset currents).

We obtain the classical algebra by deleting the central charges and the
terms in $\d q_i$ which are linear in currents and linear in $\h_i$.  These
terms have been put in square brackets.  The reader may now reconstruct
the complete $\d^\infty$ algebra by evaluating
$[\d (\e^i), \d (\e^j)], [\d (\e^\a), \d(\e^\b)]$ and $[\d (\e^i), \d)(\e^\a)]$
on $u_i$ where $\e^i = \{\e, \l^a \}$ and $\e^\a = \h^i$. We have rescaled
the results
of [4] such that the $\h$ terms in $\d_{cl} (u)$ and $\d_{cl} (v_a)$ match the
$\e$ and $\l^c$  terms in $\d_{cl} q_i$
\eq
 \int (\e^i u_A f^A{}_{i\a} \e^\a + \e^\a u_A f^A{}_{\a i} e^i ) = 0
 \eqe

 At this point we have determined all structure constants of the $\d^\infty$
algebra, as well as the transformation rules $\d^\infty u_A$.  To obtain the
transformation rules $\d^\infty h_\m{}^A$, we proceed as in the classical
case, but with $f^A{}_{BC}$ replaced by $f^A{}_{BC} + g^A{}_{BC}$, and define
\eq
\d h_\m{}^A = \del_\m \e^A + (f^A{}_{BC} + g^A{}_{BC} + 2 u_D V^{DA}{}_{BC} )
h_\m{}^C \e^B
\eqe
The origin of the factor 2 we explained before.  For $W_3$ gravity this yields
\eqa
\d h &=& \del_+ \e- h \e^\prime + h^\prime \e + (\l b^\prime - b\l^\prime
)(-4 u)
\nonumber\\
&+& (2 \l b^{\prime \prime \prime} - 3 \l^\prime b^{\prime\prime} + 3
\l^{\prime\prime} b^\prime -2 \l^{\prime\prime\prime} b) \nonumber\\
\d b &=& \e b^\prime - 2 b \e^\prime + \del_+ \l - h \l^\prime + 2 h^\prime \l
\eqae

 Direct evaluation of the commutator of two gauge transformations of $\d
h_\m{}^A$
yields
\eq
\left[ \d (\e_1), \d (\e_2) \right] h_\m{}^A  =  \d (\e_3 )h_\m{}^A
  -2  (D_\m{}^\infty u_D - m_{DE} h_\m{}^E ) V^{DA}{}_{BC} \e_1{}^C \e_2{}^B
\eqe
where $D_\m{}^\infty$ is obtained from $D_\m{}^{cl}$ by replacing
$f^A{}_{BC}$ by $f^A{}_{BC} + g^A{}_{BC}$.   This result is obtained by
using the
second and third Jacobi identity in (2).  If the expression within
parentheses vanishes
\eq D_\m{}^\infty u_D = m_{DE} h_\m{}^E \eqe
the gauge algebra closes uniformly on $u_A$ and $h_\m{}^A$.  We now shall
prove that this identity is the Ward identity for the $c \rightarrow \infty$
induced action.  We shall do so by showing that variation of this Ward
identity produces the correct $\d^\infty u_A$ transformation rules. This
argument was first given in \cite{delius}, but we repeat it here for
completeness.

Varying all gauge fields $h_\m{}^A$  in (34) into $\del_+ \e^A$, and the
leading term $\del_+ u$ into $\del_+ \d u$, one can make all terms total
$\del_+$ derivatives because the difference is terms proportional to $\del_+
u$ which can be converted into terms with $\del_-$ derivatives by using
repeatedly the Ward identity.  Extracting the overall $\del_+$, the remainder
yields indeed the correct $\d u_A$.

\section{Comments.}
We have considered classical nonlinear gauge algebras without central
terms, and obtained the $c \rightarrow \infty$ limit of the corresponding
quantum algebras, by completing the former.  This is thus a kind of Noether
procedure applied to gauge algebras.  We shall now address the question in
which cases this method works.

Consider a quantum algebra with central term
$c$, which is quadratically nonlinear.  Let it be of the generic form
\eqa
\left[ K, K \right] &=& c + K + H +\quad \frac{1}{c}: H \; H: + \ldots
\nonumber\\
\left[ K, H \right] &=& K + H \quad + \ldots \nonumber\\
\left[ H, H \right] &=& c + H + \ldots
\eqae
where the terms indicated by dots correspond to similar terms but down by
at least one factor of $c$.  For example, in the central term one may have $(c
+ 1 + \ldots)$ and in front of the $:H \; H :$ term one may have a factor
which
starts as $\frac{1}{c} +  \frac{1}{c^2} + \ldots$.  The $W_3$ algebra [6] and
the $N$ extended Bershadsky-Knizhnik superconformal algebras \cite{BK}  are
examples of this structure.  Rescale now $K = c K^\prime$
and $H = cH^\prime$, and divide by $c^2$.  Then all term are of the form
\eq
\left[ T_A^\prime , T_B^\prime \right] = \frac{1}{c} \left[ 1 + K^\prime +
H^\prime + :H^\prime H^\prime : \right] + {\cal O}\left( \frac{1}{c^2}\right)
\eqe
Moreover, the normal-ordered term $: H^\prime H^\prime :$ differs from just
the product of two $H^\prime$ by some reorderings which are of order
$\frac{1}{c}$.  Hence, to leading order in $c$, one can replace $:H^\prime
H^\prime:$ by a product, and use Poisson brackets (simple contractions)
instead of quantum commutators since further contractions are down again
by factors $1/c$.  The leading terms in $1/c$ in the Jacobi identities form
themselves an identity, and this is just the $c \rightarrow \infty$ algebra.

Next make a second rescaling in this $c \rightarrow \infty$ algebra, which
scales $H^\prime$ back to $H$, and $K^\prime$ to $K$.  This leads to an
algebra of the form
\eqa
\left[ K, K \right] &=& c + K + H + \frac{1}{c} H H \nonumber\\
\left[ K, H \right] &=& K + H \nonumber\\
\left[ H, H \right] &=& c + H
\eqae
which satisfies the classical (Poisson brackets) Jacobi identifies with
central
charges in  \cite{uscmp1}.  By further rescaling $K = \tilde{K}/\sqrt{c}$ and
subsequently taking the limit $c \rightarrow 0$, we find
\eqa
\left[ \tilde{K}, \tilde{K} \right] &=& \ H H  \nonumber\\
\left[ \tilde{K}, H \right] &=& \tilde{K} \quad , \quad [H, H] = H
\eqae
This consistent classical nonlinear algebra corresponds to the $\d_{cl}$
gauge algebra.  Hence our procedure works if the $c \rightarrow \infty$
algebra is
reductive $(f^i{}_{\a j} = 0)$, and we obtain $\d_{cl} u_\a$
by deleting the central term and all terms bilinear in $u_A$ and $\l^\b$.

It is crucial that all $[K,K]$ commutators have nonlinear $HH$ terms.
Suppose, for example, that one could subdivide the $K_\a$ generators into a
set $k_\a$ which all have nonlinear terms in the commutators among
themselves, and a set $k_a$ which do not form with $H_i$ a subalgebra.
Hence generically
\eqa
\left[ k_\a ,  k_\b \right] &=& c + K + H + \frac{1}{c} HH \nonumber\\
\left[ k_a , H \right] &=& H + k_a + k_\a
\eqae
Rescaling only $k_\a$ as $k_\a = \tilde{k}_\a / \sqrt{c}$ now produces a
singularity in the last term with $k_\a$, preventing the $c\rightarrow\infty$
limit.

Finally we explain why the gauge algebra of the transformation rules of the
fields $\varphi^i, h, b$ of $W_3$ gravity which leave the classical action
invariant, differs from the classical gauge algebra on the currents in (17).
The reason is
that invariance of the interactions $\int (h T + b W) d^2 x$ requires that $\d
h$ and $\d b$ transform contragrediently to $T$ and $W$, thus with a factor
unity in front of the $V$ term.  (The classical action is then invariant
because the
$\del_\m \e^A$ variation of $h_\m{}^A$ contribute $-u_A \del_\m \e^A$, whereas
the $\varphi$-kinetic term varies into $-(\del_\m u_A)\e^A$).  On the other
hand,
uniform closure on $h_\m{}^A$ and $u_A$ requires a factor 2 in front of the
$V$ term
in $\d h$, and this explains why the $\l$ commutator on $b$ in (15) is a
factor 2
smaller than the $\l$ commutator on $u$ and $v$ in (17).

In [9] we considered a different approach to the gauging of nonlinear
algebras, in
which we introduced in addition to $h_\m^A$ further {\it scalars} $t_A$
instead of
currents $u_A{}^\m$.  These scalars played the role of Higgs scalars in a $d=4$
Yang-mills model.  Whether scalars or currents, one can also view them as
auxiliary
fields which close a gauge algebra.

 As was shown in this paper many properties of Lie algebras carry over to the
case of non-linearly generated Lie algebras. However, two main features which
are well understood in the case of linear Lie algebras are not at all
established in the case of non-linear algebras. The first is how to tensor
representations such as to obtain new representations. This problem is
obviously directly related to the presence of non-linearities in the
commutation relations. The second and probably most important problem is the
lack of understanding of the geometry behind non-linearly generated Lie
algebras.

\noindent {\bf Acknowledgements:}  The work of A. Sevrin  was
supported in part by the Director Office of Energy Research, Office of High
Energy and Nuclear Physics, Division of High Energy Physics of the U.S.
Department of Energy under Contract DE-AC03-765F00098, and in part by the
National Science Foundation under grant PHY90-21139.  The work of P. van
Nieuwenhuizen was supported in part by NSF grant PHY92-11367.

\end{document}